\documentclass[aps,
prb,
reprint,
superscriptaddress,
amsmath,amssymb,
floatfix,
raggedbottom,
]{revtex4-2}

\newcommand{\orcid}[1]{\href{https://orcid.org/#1}{\includegraphics[width=8pt]{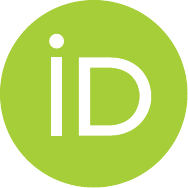}}}

\usepackage[T1]{fontenc}
\usepackage[utf8]{inputenc}

\usepackage[breaklinks=true,colorlinks=true,linkcolor=blue,urlcolor=blue,citecolor=blue]{hyperref}
\usepackage{graphicx}

\usepackage{braket}
\usepackage{lipsum}
\usepackage{siunitx}
\usepackage{soul}

\usepackage{chemformula}
\graphicspath{ {figures/} }

\begin{document}

\title{Disorder effects of vacancies on the electronic transport properties of realistic topological insulators nanoribbons: the case of bismuthene}

\author{Armando Pezo \orcid{0000-0003-2613-3808}}\email{armando.pezo@ufabc.edu.br}
\affiliation{Federal University of ABC (UFABC), 09210-580, Santo André, São Paulo, Brazil}
\affiliation{Brazilian Nanotechnology National Laboratory (LNNano), CNPEM, 13083-970, Campinas, São Paulo, Brazil}

\author{Bruno Focassio \orcid{0000-0003-4811-7729}}
\affiliation{Federal University of ABC (UFABC), 09210-580, Santo André, São Paulo, Brazil}
\affiliation{Brazilian Nanotechnology National Laboratory (LNNano), CNPEM, 13083-970, Campinas, São Paulo, Brazil}

\author{Gabriel R. Schleder \orcid{0000-0003-3129-8682}}
\affiliation{Federal University of ABC (UFABC), 09210-580, Santo André, São Paulo, Brazil}
\affiliation{Brazilian Nanotechnology National Laboratory (LNNano), CNPEM, 13083-970, Campinas, São Paulo, Brazil}

\author{Marcio Costa \orcid{0000-0003-1029-8202}}
\affiliation{Instituto de F\'{\i}sica, Universidade Federal Fluminense, 24210-346  Niter\'oi, Rio de Janeiro, Brazil}
\affiliation{Brazilian Nanotechnology National Laboratory (LNNano), CNPEM, 13083-970, Campinas, São Paulo, Brazil}

\author{Caio Lewenkopf \orcid{0000-0002-2053-2798}}
\affiliation{Instituto de F\'{\i}sica, Universidade Federal Fluminense, 24210-346  Niter\'oi, Rio de Janeiro, Brazil}

\author{Adalberto Fazzio \orcid{0000-0001-5384-7676}}\email{adalberto.fazzio@lnnano.cnpem.br}
\affiliation{Brazilian Nanotechnology National Laboratory (LNNano), CNPEM, 13083-970, Campinas, São Paulo, Brazil}
\affiliation{Federal University of ABC (UFABC), 09210-580, Santo André, São Paulo, Brazil}

\date{\today}

\begin{abstract}
The robustness of topological materials against disorder and defects is presumed but has not been demonstrated explicitly in realistic systems. In this work, we use state-of-the-art density functional theory and recursive nonequilibrium Green’s functions methods to study the effect of disorder on the electronic transport of long nanoribbons, up to $157\;\rm nm$, as a function of vacancy concentration. In narrow nanoribbons, a finite-size effect gives rise to hybridization between the edge states erasing topological protection. Hence, even small vacancy concentrations enable backscattering events. We show that the topological protection is more robust for wide nanoribbons, but surprisingly it breaks down at moderate structural disorder. Our study helps to establish some bounds on defective bismuthene nanoribbons as promising candidates for spintronic applications.
\end{abstract}

\maketitle

\section{Introduction}

Topological materials have been intensively studied in recent years \cite{Hasan2010,Moore2010a,Qi2011,ando_review,Bansil2016,Giustino2020} unveiling interesting new physics and opening new applications possibilities in spin-based electronic devices \cite{RevModPhys.76.323}. 
Of particular interest are large band gap topological insulators (TIs), that are good candidates for the realization of the quantum spin Hall (QSH) effect at room temperature \cite{Reis2017, Marrazo2019}.
In two-dimensional (2D) QSH insulators, the edges of the sample carry metallic states that are protected by time-reversal symmetry (TRS)~\cite{kane_mele_qshe_graphene,z2_KaneMele.95.146802,Bernevig2006, Konig2007, PhysRevB.81.235323, NJP2010, PhysRevB.83.081402} and decay exponentially into the bulk \cite{Drozdov2014,finite_size_effects,depth_loca_k_dependency}. Moreover, theory predicts that in nanoribbons these edge states carry dissipationless helical spin currents. The penetration depth, that quantifies the edge states exponential decay rate, is (roughly) inversely proportional to the band gap \cite{finite_size_effects,Shen2011} and plays a key role in the nanoribbon transport properties. As it happens with the length scales of 3D topological insulators \cite{Zhang_2010}, to fully display the features of a TI, the nanoribbon width must be much wider than the penetration depth $\xi$ of the edge states, otherwise they can easily hybridize.  

Among several candidates for topological materials, bismuthene, also known as buckled or bilayer bismuthene, is of special interest due to its large electronic band gap of \SI{0.5}{\electronvolt}, along with its structural stability and large spin-orbit coupling (SOC) \cite{Aktrk2016,Kadioglu2017,Wang2017_1}. The existence of charge puddles and other types of defects that could be detrimental for the formation of a topological phase do not play an important role in bismuthene, as experimentally demonstrated \cite{Reis2017}. Indeed, since its experimental realization, it was proposed that even amorphous structures of bismuthene occurring before the annealing process \cite{Costa2019,focassio2020amorphous}, as well as strongly disordered systems \cite{Ni2020}, support topological states. The robustness of bismuthene non-trivial topology makes it ideal for material design, both by tuning the lattice constant and effective spin-orbit coupling (SOC) by epitaxial constraint or by substitutional alloying with lighter elements, such as Sb or As, without causing a topological transition~\cite{Wang2017_2,Aktrk2016}. The presence of non-magnetic defects leaves the band topology unchanged in these materials \cite{Wang2017_1,Costa2019,focassio2020amorphous,Ni2020}.

Two recent works have proposed that vacancies can spoil the topological (elastic) transport properties of 2D TIs. Studying a generic tight-binding model with a Hubbard mean-field term, Ref.~\cite{Novelli2019} has shown that vacancy induced localized states can give rise to local magnetic moments and destroy the topological protection. In turn, Ref.~\cite{Tiwari_2019} proposes a very different mechanism. Based on another tight-binding toy model, Ref.~\cite{Tiwari_2019} has numerically shown that small concentrations of vacancy defects do not eliminate the topological edge states, but can cause inter-edge state hybridization in certain energies intervals within the topological gap \cite{Tiwari_2019}. For the BHZ model \cite{Bernevig2006}, several works have addressed the role of strong disorder, demonstrating a modification of the local currents and inter-edge tunneling effects \cite{Lu2011,Chu2012,Lee2013,Dang2015}. The dominant mechanism depends critically on the choice of the schematic-model Hamiltonian, which---from the point of view of materials---is very unsatisfactory. To better 
understand the interplay between edge states and defects a material-specific systematic study using \textit{ab initio} methods is in order.

Vacancies on bismuthene show small formation energies where $sp^2$ bismuth dangling bonds can be distributed around the vacancy leading to resonances in the band gap, modifying the electronic properties of the host material. Theoretical calculations on bismuthene show no evidence of magnetic moments induced from vacancies~\cite{Kadioglu2017}, preserving TRS and retaining its non-trivial topological band structure. Recent \textit{ab initio} calculations demonstrate that transport along the edges is insensitive to vacancies created in ultra-narrow TI zigzag nanoribbons \cite{vannucci2020conductance}. Even though these vacancies allow the development of magnetic moments, the perfect conductance for energies within the bulk gap is recovered provided the vacancies are passivated by hydrogen. 
Our study, in turn, addresses more realistic system sizes, namely, both much wider and much longer, unveiling a different kind of disordered-induced transport mechanism. 

In this work, we investigate the electronic transport properties of buckled bismuthene nanoribbons of realistic sizes using the full Hamiltonian in the orbital representation obtained from density functional theory (DFT)~\cite{dft1964,dft1965,review} calculations combined with recursive nonequilibrium Green’s functions (NEGFs)~\cite{Caroli_1971,trace_formula,Datta1995,transampa}. 
First, we explore the backscattering mechanism due to inter-edge hopping mediated by vacancy localized states in different nanoribbon widths and discuss the detrimental effects in transport properties caused by single vacancies in narrow nanoribbons. 
We proceed to investigate the robustness of the topological properties by studying the conductance of ribbons as a function of their widths, lengths (up to $ 157\;\rm nm$), and vacancy concentration. We show that narrow ribbons, whose widths $w$ are comparable with the edge states penetration depth $\xi$, present the onset of Anderson localization effects already at low vacancy concentrations. 
In distinction, for wide nanoribbons, where $w \gg \xi$, the edge states maintain a quantized conductance in the topological gap at low disorder concentrations. Surprisingly, we find that topological protection is destroyed already at modest vacancy concentrations.

 \section{Computational Methods}\label{sec:methods}

Our calculations combine the flexibility of a plane-wave basis set to obtain the optimized structures with a localized basis for electronic transport. In both cases, we use the Perdew-Burke-Ernzerhof (PBE)~\cite{gga,pbe} exchange-correlation functional. We perform the geometry optimizations with a plane-wave basis as implemented in the \textsc{VASP} package~\cite{vasp1,vasp2}. 
In these calculations, we employ \SI{400}{\electronvolt} for the plane-wave expansion cutoff and ionic potentials are described using the projector augmented-wave (PAW) method~\cite{paw}. 
Vacancies are modeled by removing an atom from the bulk and performing the geometry optimization with force criterion of \SI{5e-3}{\electronvolt\per\angstrom}.

The transport calculations use full \textit{ab initio} DFT Hamiltonian matrices obtained directly from the \textsc{SIESTA} code~\cite{siesta_method}, employing atom-centered single-$\zeta$ plus polarization (SZP) basis sets.
We use the energy cutoff for real-space mesh of \SI{350}{Ry}, sampling the reciprocal space with 10 $\mathbf{k}$-points along the periodic direction of the presented nanoribbons, which edges were hydrogen-passivated.
We add \SI{20}{\angstrom} of vacuum in both non-periodic directions to avoid spurious interactions between periodic images. The self-consistent SOC is introduced via an on-site approximation~\cite{siesta_on-site_soc} using fully-relativistic norm-conserving pseudopotentials \cite{tm_pseudopotentials}. The system Hamiltonian and overlap matrices are obtained after performing a full self-consistent cycle.

\begin{figure}[!t]
    \centering
    \includegraphics[width=\linewidth]{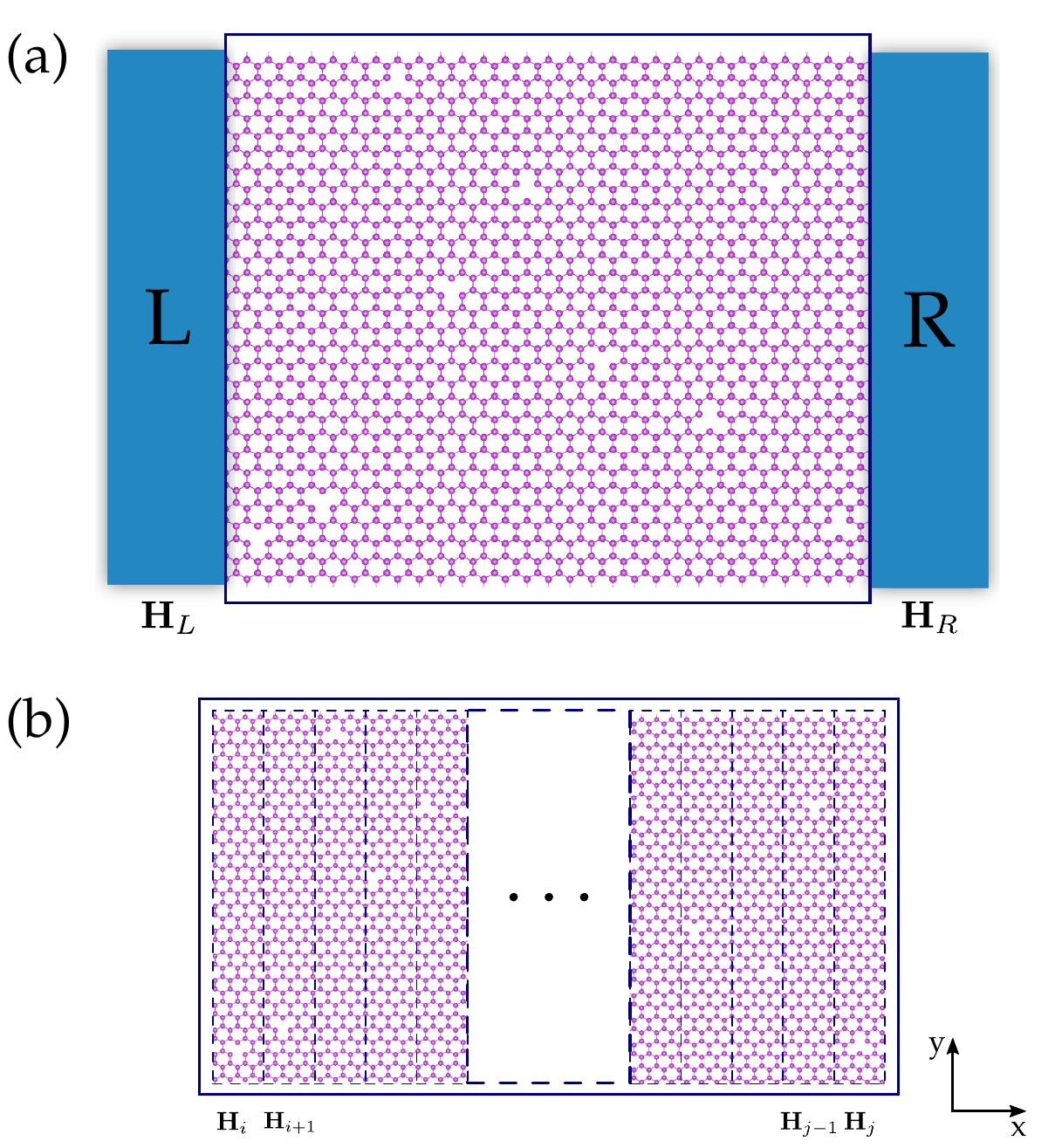}
    \caption{Schematic representation of the two-probe setup studied in this work. (a) The electrodes are located at the left (L) and right (R) sides of the device. (b) Zoomed-in view of the scattering region (S), illustrating some of its building blocks containing vacancies.}
    \label{fig:schema_dec}
\end{figure}

The electronic transport calculations are implemented following the standard NEGF approach \cite{Caroli_1971,trace_formula,Datta1995}. We consider a two-probe terminal setting, as illustrated in Fig.~\ref{fig:schema_dec}(a). The left (L) and right (R) electrodes are modeled by semi-infinite pristine zigzag bismuthene leads. 

Our approach employs a decimation technique \cite{tb-green,rossi_transport,transport-carbon-nano,deci_james,Lewenkopf2013,2053-1583-2-2-022001} that allows us to address large system sizes. The scattering region is partitioned into building blocks connected by first neighbour interactions. Each building block is computed through DFT. The procedure takes into account all degrees of freedom of the system comprised of all the building blocks. The partition scheme is  depicted in Fig.~\ref{fig:schema_dec}(b). Accordingly, the system Hamiltonian is expressed in the localized basis by the block matrix 
\begin{equation}
\mathbf{H} = \begin{pmatrix}
\mathbf{H}_L & \mathbf{H}_C & 0  \\
\mathbf{H}^\dagger_C & \mathbf{H}_S & \mathbf{H}_C \\
 0 & \mathbf{H}^\dagger_C & \mathbf{H}_R \\
\end{pmatrix}\label{eq:hamiltonian_two_lead_setup}
\end{equation}
where $\mathbf{H}_L$ and $\mathbf{H}_R$ are the Hamiltonian matrices describing the left and right electrodes, while $\mathbf{H}_C$ is the coupling between the leads and the central region, and $\mathbf{H}_S$ is the tridiagonal block matrix 
\begin{equation}
\mathbf{H}_S = \begin{pmatrix}
\mathbf{H}_1 & \mathbf{H}_C & 0 & \hdots & 0 \\
\mathbf{H}^\dagger_C & \mathbf{H}_2 & \hdots & \hdots &\vdots \\
 0 & \vdots & \ddots &\vdots &0 \\
 \vdots & \hdots & \hdots &  \mathbf{H}_{N-1} &\mathbf{H}_C \\
0 & \hdots & 0 & \mathbf{H}^\dagger_C & \mathbf{H}_N \\
\end{pmatrix}\label{eq:big_hamiltonian}
\end{equation}
representing the scattering region (S). We take $\mathbf{H}_C$ also as the coupling between each building block. We consider a uniform distribution of vacancies within the building blocks represented in Fig. \ref{fig:schema_dec}(b), except for a small region of \SI{2.175}{\angstrom} at the simulation box boundaries in the periodic direction. This is necessary to properly connect successive building blocks by $\mathbf{H}_C$ as in Eq.~\eqref{eq:big_hamiltonian}.

The scattering region spin resolved retarded Green's function~\cite{Datta1995,data-negf} reads
\begin{equation}
    \mathbf{G}^{r}_{S}(E)= \left( E^+ \mathbf{S}_{S} -\mathbf{H}_{S} -\mathbf{\Sigma}_L^{r} - \mathbf{\Sigma}_R^{r} \right)^{-1}
\end{equation}
where $E^+ = \lim_{\delta \to 0^+} E+i\delta$, $\mathbf{S}_S$ is the overlap matrix, and $\mathbf{\Sigma}^{r}_{R/L}(E)=(E^+ \mathbf{S}_C-\mathbf{H}_C)\mathbf{G}^{r}_{0,R/L}(E^+\mathbf{S}^{\dagger}_C-\mathbf{H}_C^{\dagger})$ are the embedding self-energies that account for the system decay width due to the coupling with the leads. Here $G^{r}_{0,R/L}$ is the retarded surface Green's function of the $R/L$ electrode~\cite{Sancho_1985,tb-green}.
The block structure of $\mathbf{H}_S$ allows for a  very efficient computation of $G^r_S(E)$ using the recursive Green's function method (see, for instance, Ref.~\cite{Lewenkopf2013}, for a review).

In the linear response, at small bias, the zero-temperature conductance is given by the Landauer formula $G(E_F) = (e^2/h)\;T(E_F)$, where the transmission $T$ reads \cite{Caroli_1971,trace_formula,Datta1995}
\begin{equation}
    T(E)=\text{Tr}\left[\mathbf{\Gamma}_L(E) \mathbf{G}_{S}^{a} (E)\mathbf{\Gamma}_R(E) \mathbf{G}_{S}^{r}(E)\right]
\end{equation}
where $\mathbf{G}_{S}^{a} = [\mathbf{G}_{S}^{r}]^\dagger$ and the decay width matrices $\mathbf{\Gamma}_{L/R}$ are given by  $\mathbf{\Gamma}_{L/R}={i}(\mathbf{\Sigma}_{L/R}^r-\mathbf{\Sigma}_{L/R}^{a})$.

 \section{Results and Discussion}
\label{sec:resutls}

In this section we analyze the effect of vacancies on the transport properties of bismuthene zigzag nanoribbons. We begin by discussing the topological properties of pristine nanoribbons of different representative widths. Next, we investigate the effect of a single-vacancy on the electronic and transport properties of these systems. Finally, we study the conductance of these systems for different vacancy concentrations and nanoribbon widths.

\begin{figure}[!htb]
    \centering
    \includegraphics[width=.85\linewidth]{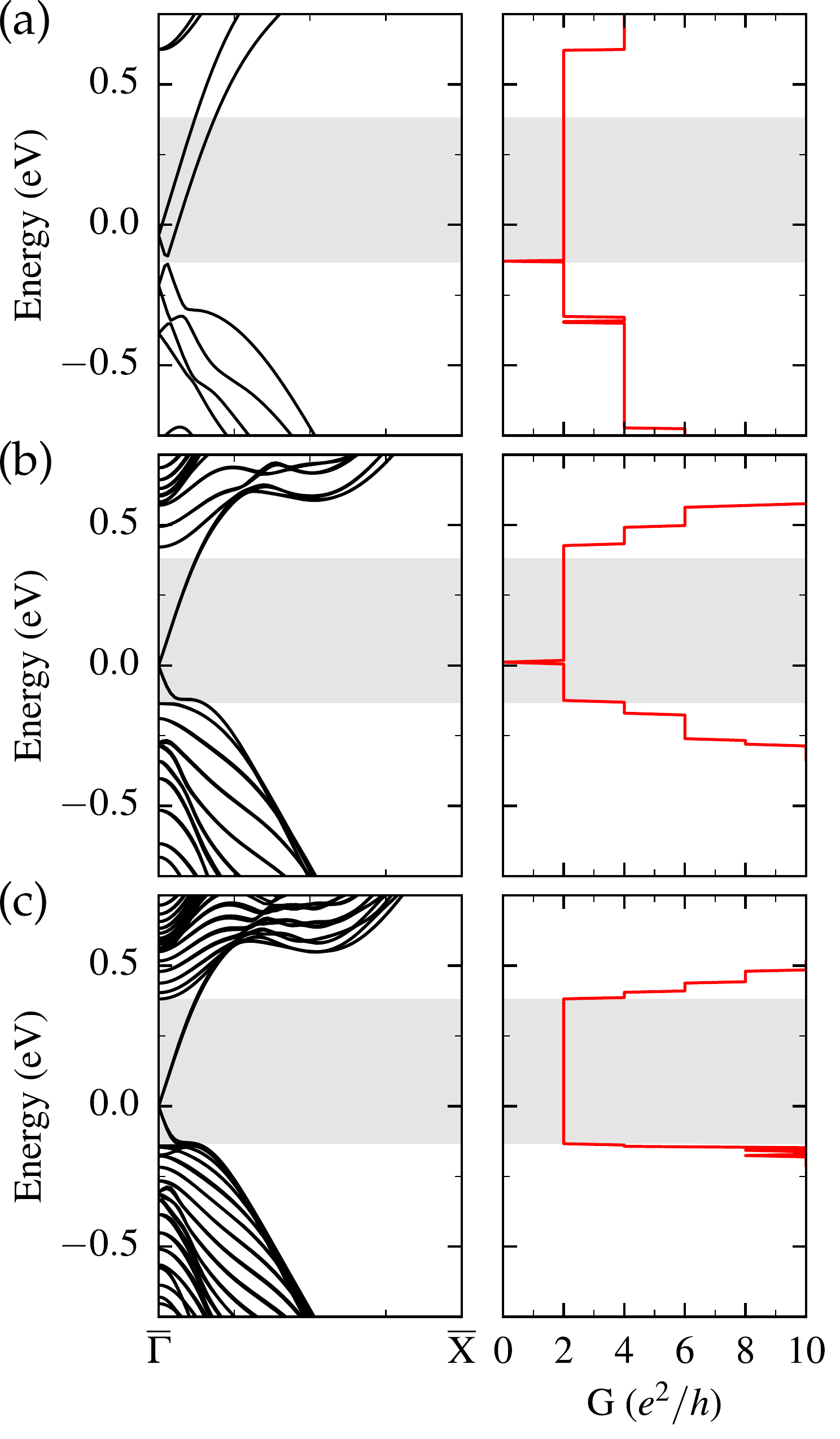}
    \caption{Electronic band structure along the $\overline{\Gamma}-\overline{X}$ direction (left panels) and the corresponding conductance $G$ (right panels) for pristine (infinite length) bismuthene nanoribbons of widths (a) $w_{20}=$ \SI{20}{\angstrom}, (b) $w_{65}=$ \SI{65}{\angstrom}, and (c) $w_{110}=$ \SI{110}{\angstrom}. The gray energy windows correspond to the bulk topological gap. 
    }
    \label{fig:bands_pristine}
\end{figure}

Let us first investigate the effect of the ribbon width on pristine systems. Due to translation invariance these systems can be viewed as having infinite length.
In Fig.~\ref{fig:bands_pristine} we show the electronic band structure and the conductance for bismuthene nanoribbons with 3 different representative widths. To help the discussion, the bulk-topological gap $\Delta_{\rm TG}$, corresponding to $ \SI{-0.1}{\electronvolt} < E < \SI{0.4}{\electronvolt}$, is indicated in grey.
Figure ~\ref{fig:bands_pristine}(a) shows the results for a narrow nanoribbon, $w_{20}$  of width \SI{20}{\angstrom}, comparable to those of ref.~\cite{vannucci2020conductance}. 
The electronic states bridging the bulk-topological $\Delta_{\rm TG}$ are split and the system displays a small gap, in distinction to the standard picture of topological protected metallic edge states in very large systems.
Figure ~\ref{fig:bands_pristine}(b) shows an intermediate width nanoribbon, namely $w_{65}$, of \SI{65}{\angstrom} width. Here, there is still a small gap at the $\Gamma$-point, though the edge states are degenerate as expected for a topological insulator. In addition, the top of the valence and bottom of the conduction trivial state bands approach the corresponding bulk energies, indicating that finite width quantization effects are much smaller than in the previous case. 
Finally, Fig.~\ref{fig:bands_pristine}(c) shows a wide nanoribbon, $w_{110}$  of width \SI{110}{\angstrom}, with no gap and  degenerate edge states with helical texture, as expected for a TI. Here, the bulk-topological gap corresponds very closely to the energy interval where one finds only edge states. In all cases, the conductance $G(E)$ is quantized and the conductance steps are observed as expected for pristine systems~\cite{Datta1995}. 

The lack of topological protection in narrow ribbons is a result of strong overlap between the states at the opposite system edges. This can be understood in terms of the spatial localization of edge states or, more precisely, their so-called penetration length $\xi$. The latter can be roughly estimated from the mass term in the $k \cdot p$ Dirac Hamiltonian  describing the inverted bulk band gap as \cite{Shen2011,shen2017topological, Edge_physics,C4CP02213K}
\begin{equation}
\xi \approx \hbar v_F/\Delta_{\rm TG},\label{eq:penetration_depth}
\end{equation}
where $v_F$ is the Fermi velocity corresponding to the topological bands, namely, $\hbar v_F = d\varepsilon_{k}/dk$.  
By estimating the Fermi velocities from  Figs.~\ref{fig:bands_pristine}(b) and (c) we find: For the $w_{65}$ nanoribbon $v_F =$  \SI{4.58e5}{\metre\per\second}, that renders
$\xi = 0.6\;\rm nm$, while for $w_{110}$ the Fermi velocity is \SI{5.79e5}{\metre\per\second} and 
$\xi = 0.8\;\rm nm$. 
These estimates of $\xi$ for buckled bismuthene are slighlty larger than the values reported in experiments
\cite{fermi_velo_depth,Reis2017} that obtain $\xi \approx 0.4\;\rm nm$ in SiC(0001) supported flat bismuthene.
We stress that these are different  material systems. The discrepancy can be explained by recalling that $\Delta_{\rm TG}$ of the buckled bismuthene is smaller than the band gap of the flat one. 

\begin{figure}
    \centering
    \includegraphics[width=\linewidth]{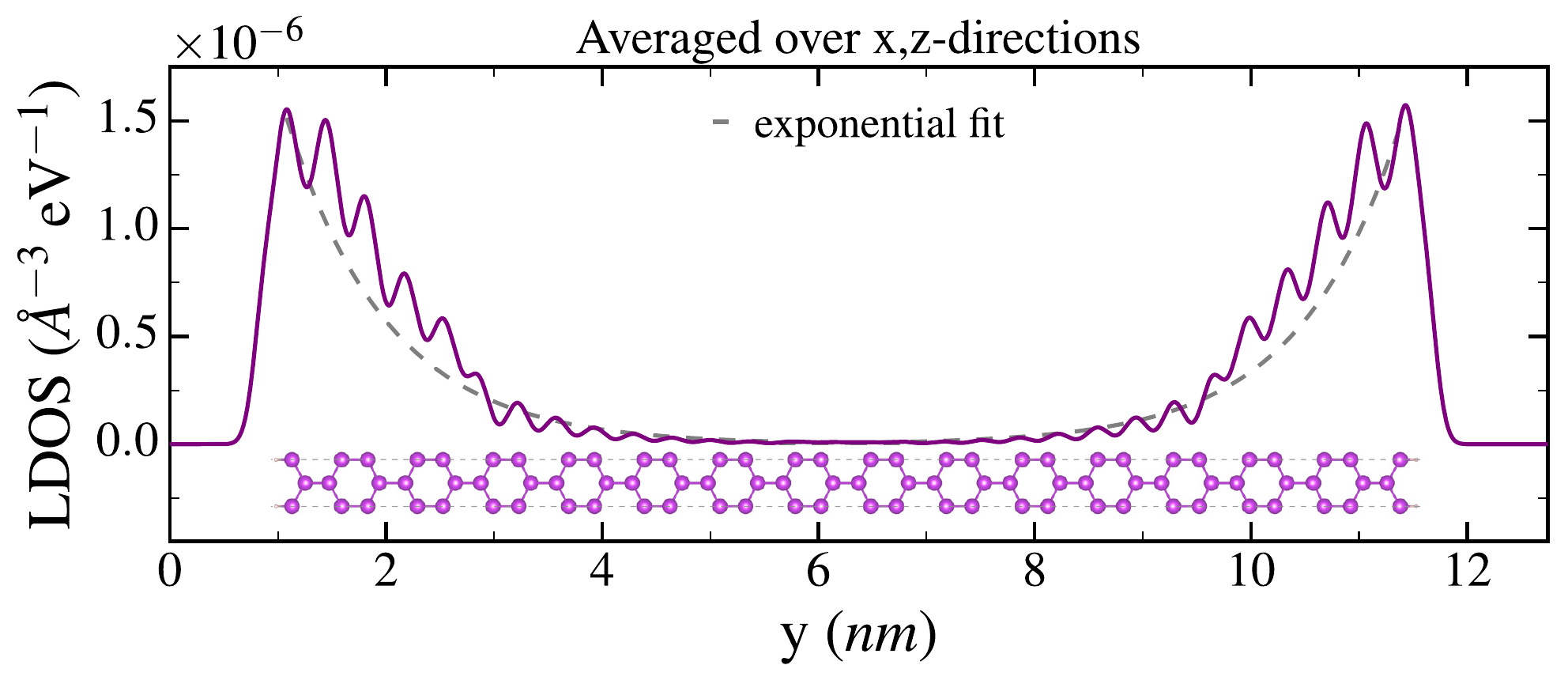}
    \caption{Local density of states (LDOS) at the Fermi level ($E_F = 0$) for pristine (infinite length) $w_{110}$ nanoribbon. The LDOS is averaged over the $x$ and $z$ directions and the y-axis corresponds to the transverse direction along the ribbon width. The gray dashed line shows the exponential fit. The geometry below the LDOS shows the bismuthene $w_{110}$ nanoribbon view perpendicular to the $x$--$y$ plane.}
    \label{fig:penetration}
\end{figure}

A more quantitative estimate of $\xi$ is taken from the local density of states (LDOS) averaged over the $x,z$ directions as a function nanoribbon transversal axis $y$, see Fig.~\ref{fig:penetration}. Fitting an exponential function, we estimate the penetration length as $\xi = 0.9 \rm\;nm$ for the $w_{110}$ ribbon, in good agreement with $\xi$ 
obtained using Eq.~\eqref{eq:penetration_depth}. Using $\xi$, we can also estimate a threshold length for the LDOS decay for which the interedge states hybridization becomes negligible. For instance, at distances $\Delta y \simeq 2.8\;\rm nm$ from the edges, the LDOS decays by $\sim 95\%$. Hence, we expect that narrow nanoribbons with $w \alt 2 \Delta y$ lack topological protection due to the strong overlap of states localized at opposite system edges. The $w_{65}$ nanoribbons are at the crossover between non-protected and topologically protected phases.

Next, we investigate ribbons containing a single-vacancy. For a slab geometry, we calculate the formation energy $E_V$ of these single vacancies systems using the following expression \cite{C5RA23052G}
\begin{equation}
    E_V = E_{\rm tot}-(E_{\rm pristine}+\mu_{V} N_V )
\end{equation}
where $E_{\rm tot}$ is the total energy of the single-vacancy system given by a fully relaxed DFT calculation, $E_{\rm pristine}$ is the energy for the pristine ribbon, $N_V$ is the number of vacancies (in our case $N_V =1$), and $\mu_{V}$ is the chemical potential to remove a bismuth atom. Here, we take $\mu_{V}$ as the energy per atom of the pristine bismuthene monolayer. 
For purposes of comparison, we simulate a $5 \times 5$ supercell containing a vacancy. By doing so, we avoid that theses vacancies interact with their periodic images obtaining a formation energy of \SI{1.04}{\electronvolt}, which is in agreement with previous reports \cite{Kadioglu2017}. 

\begin{figure}[!htb]
    \centering
    \includegraphics[width=\linewidth]{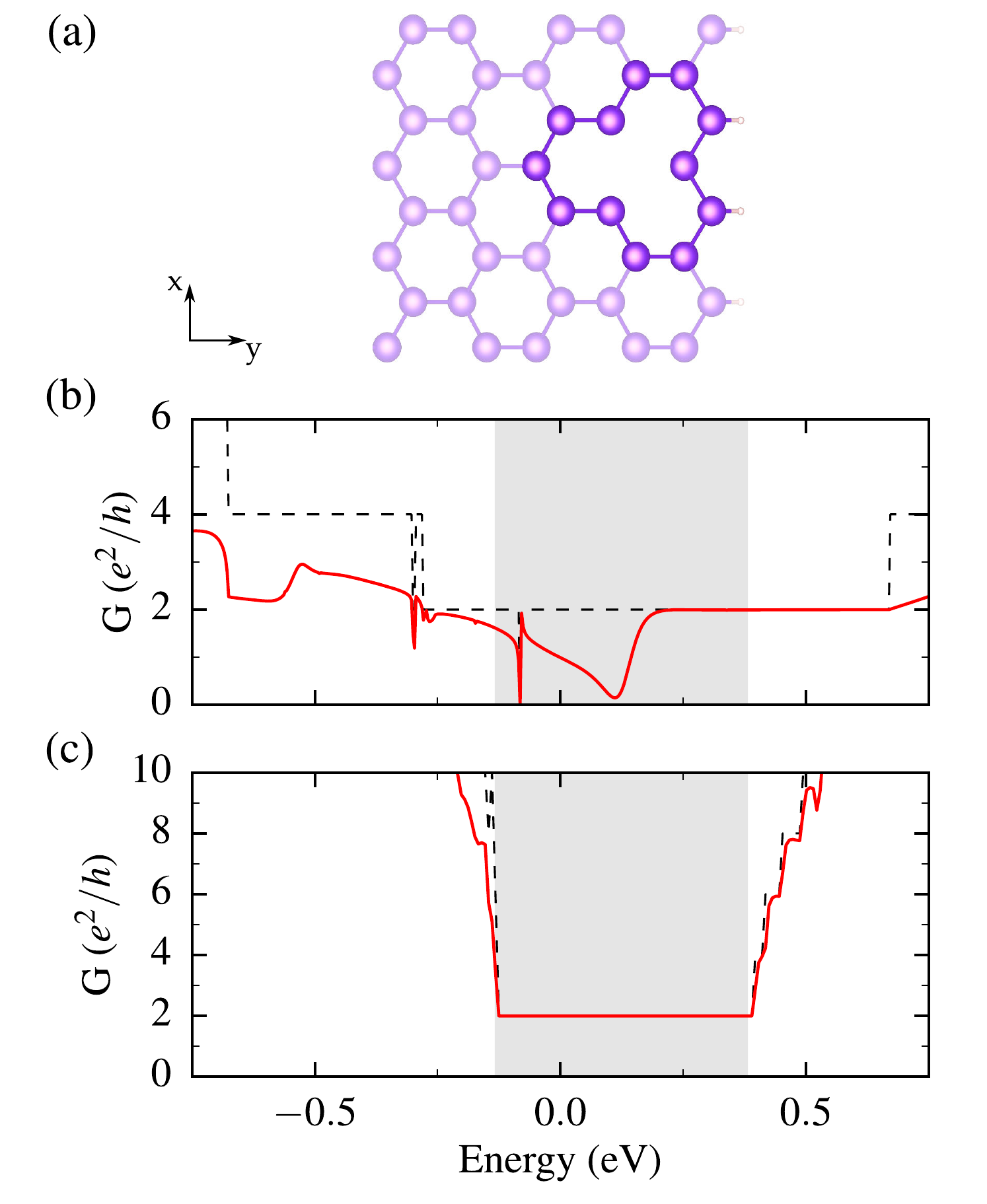}
    \caption{Single-vacancy close to the edge of a bismuthene nanoribbon: (a) Lattice structure and conductance for (b) $w_{20}$ nanoribbon, and (c) $w_{110}$ nanoribbon. The gray region indicates the bulk topological gap. The black dashed line corresponds to the conductance in the pristine case. All structures consist of a single building block $N=1$ containing the vacancy coupled to semi-infinite pristine ribbons.}\label{fig:bands_conductance_vacancy}
\end{figure}

Figure \ref{fig:bands_conductance_vacancy} shows the conductance of the  $w_{20}$ and $w_{110}$ ribbons in the presence of a single vacancy placed close to one of the system edges. For this calculation, the scattering region corresponds to a single building block $N=1$ containing the vacancy.
For the $w_{20}$ ribbon the quantized conductance is destroyed, showing that the edge states are not robust against disorder. 
In turn, the $w_{110}$ ribbon shows no deviation from perfect conductance $G_0 = 2e^2/h$ within the topological gap. 

Several studies on a variety of 2D materials indicate that single-vacancies give rise to localized states \cite{Yazyev2010,Novelli2019}. It has been further shown that such disorder-induced localized states cause the formation of local magnetic moments \cite{Palacios2008,Ugeda2010,Yazyev2010,Miranda2016,Miranda2019} that, if close to the system edges, can be detrimental for the topological protection, differently from dual topological insulators \cite{Focassio2020}. 
Let us study the relevance of these findings to bismuthene.

\begin{figure}[!htb]
	\centering
	\includegraphics[width=\linewidth]{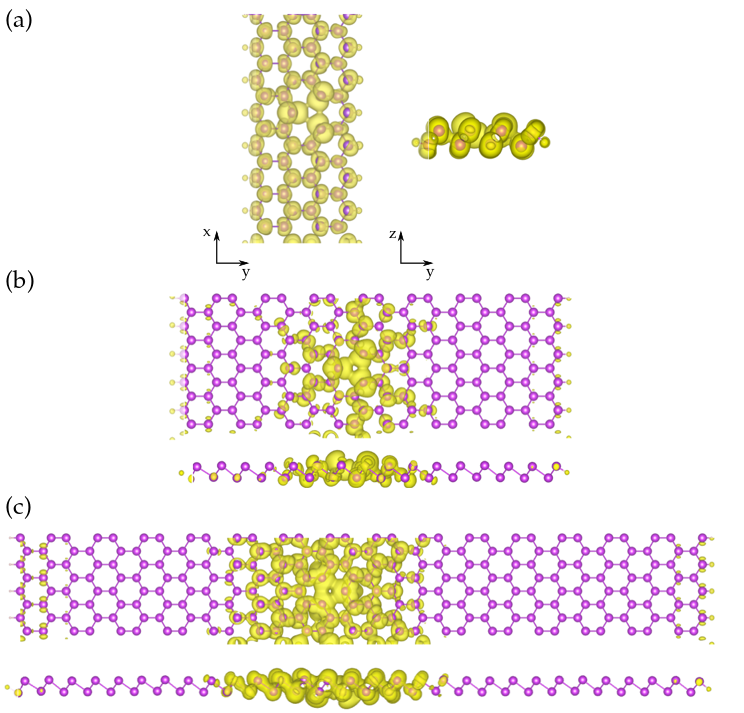}
	\caption{Top ($x$--$y$ axis) and lateral ($y$--$z$ axis) projections of the local density of states (LDOS) for single vacancy in (a) $w_{20}$, (b) $w_{65}$ and (c) $w_{110}$ nanoribbons. The isosurface value for the LDOS is of \SI{0.025}{\per\cubic\angstrom\per\electronvolt}, corresponding to an energy range $\SI{0.2}{\electronvolt} < E < \SI{0.3}{\electronvolt}$. All structures consist of a single building block $N=1$ containing the vacancy coupled to semi-infinite pristine ribbons.}
	\label{fig:ldos}
\end{figure}

Figure \ref{fig:ldos} shows the local density of states (LDOS) for the $w_{20}$, $w_{65}$ and $w_{110}$ nanoribbons calculated for the energy window $\SI{0.2}{\electronvolt} < E < \SI{0.3}{\electronvolt}$. The strong enhancement of the LDOS centered around the vacancy corresponds to exponentially decaying orbitals paired after the atom relaxation. Figure \ref{fig:ldos}(a) shows that the vacancy states increase the overlap between inter-edge states, whereas for the $w_{65}$ nanoribbon the overlap is small even in the presence of a vacancy. For the $w_{110}$ ribbon shown in Fig. \ref{fig:ldos}(c) the overlap between the edge states and the vacancy localized states is negligible.

Our \textit{ab initio} fully relativistic calculations show that, regardless of the superlattice size, type of basis-set used in the calculation, and position of the vacancy (bulk or nanoribbon edge), there is no indication of a defect-induced magnetic moment, more precisely, we find $\mu< 10^{-4} \rm\mu_B$. Interestingly, when the SOC is artificially turned off, our spin polarized  calculations give a magnetic moment of $0.8\;\rm\mu_B$ around the vacancy. This suggests that the local orbital hybridization, mainly $s$ and $p$, caused by the spin-orbit interaction is probably responsible for the spin-polarization quench in bismuthene. Besides, the low energy electronic properties of bismuthene are dominated by $p$-orbitals, and in this material, there is no evidence of strong interaction effects due to localized electrons; therefore the non-magnetic ground state is adequately described by DFT \cite{Malyi2020,Zunger2020,Kadioglu2017,Drozdov2014}. Based on these results, we rule out vacancy induced magnetic moments as a mechanism \cite{Novelli2019} to hinder the topological protection in bismuthene.

We now study the conductance $G$ as a function of vacancy concentration. This is done by modelling nanoribbons as a sequence of $N$ building blocks with a given concentration $n_{\rm V}$ of randomly placed vacancies, as described in Sec.~\ref{sec:methods}. The calculation of the electronic properties of the individual building blocks is the main computational bottleneck of our study. Smaller building blocks optimize the computation time, but their lengths $\ell$ have to be large enough to correctly describe the vacancy-induced localized states. We find that $\ell = \SI{17.4}{\angstrom}$ gives accurate results with a feasible computational time 
even for the wider ribbon. We choose $N=90$ to address nanoribbons of realistic sizes. Here, the considered scattering region has a total length of \SI{156.6}{\nano\metre}.

\begin{figure*}
\centering
\includegraphics[width=\linewidth]{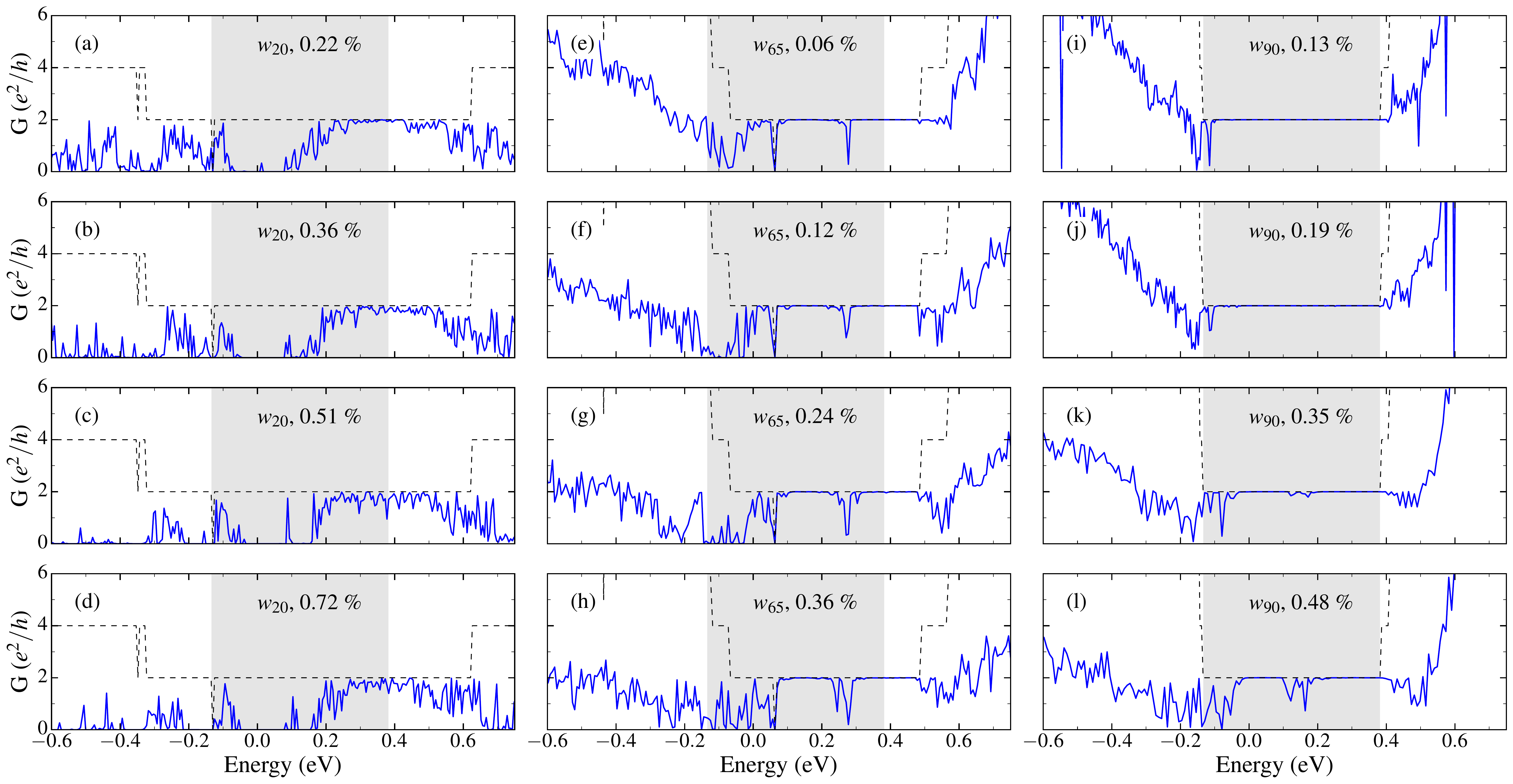}
\caption{Conductance $G$ (in units of $e^2/h$) as a function of the energy $E$ (in eV) for several vacancy concentrations for the $w_{20}$, $w_{65}$, and $w_{110}$ nanoribbons. The ribbon widths and vacancy concentrations are indicated in each panel. The blue line corresponds to $G$ in the presence of disorder, while the black dashed one stands for $G$ in the pristine case. The gray region indicates the topological gap. In all cases the length is \SI{157}{nm}.
}
\label{all_transm}
\end{figure*}

Figure~\ref{all_transm} shows that vacancies, even in small concentrations, have a strong effect on the transport properties of bismuthene nanoribbons. 
In all cases, the pristine conductance quantization is destroyed outside the topological gap $\Delta_{\rm TG}$. 
For instance, for $n_{\rm V}\approx 0.2\%$, one observes the onset of localization ($G/G_0 \ll 1$) for $w_{20}$ and a very strong suppression of $G$ for $w_{110}$, when compared to the pristine case.
For energies outside $\Delta_{\rm TG}$, bismuthene is an ordinary material and the results can be interpreted in terms of standard disorder-induced backscattering mechanisms. In what follows we discuss the behavior of $G$ for energies within the topological phase.

Figures \ref{all_transm}(a) to \ref{all_transm}(d) correspond to the narrow width nanoribbon $w_{20}$ with vacancy concentration $n_V$ ranging from  $0.22\;\%$ to $0.72\;\%$. As discussed above, these systems are not topological insulators. 
Interestingly, for increasing $n_V$ the $w_{20}$ ribbons behave as trivial insulators for energies outside the topological gap, but the conductance is less 
suppressed inside $\Delta_{\rm TG}$. 
A similar behavior has been studied in graphene \cite{Wakabayshi2007,Lima2012} and MoS$_2$ nanoribbons \cite{Ridolfi2017,Pezo2019}. 
The large momentum transfer necessary to enable backscattering processes, as inferred from the trivial pristine nanoribbon electronic band structure in Fig. \ref{fig:bands_pristine}(a), preserves the conductance in narrow energy intervals.
See, for instance \cite{Wakabayshi2007,Lima2012} for detailed discussions.

Figures \ref{all_transm}(e) to \ref{all_transm}(h) correspond to $w_{65}$ nanoribbons of intermediate width and $n_V$ ranging from $0.06\,\%$ to $0.36\,\%$. 
For energies outside $\Delta_{\rm TG}$ the conductance is strongly suppressed with respect to the pristine case, as expected for a trivial disordered system. In contrast to the $w_{20}$ case, for $n_V \alt 0.36\,\%$ the system does not display any Anderson localization features. More interestingly, the conductance is close to $G_0$ within the topological bandgap. The edge states are not fully protected by topology: there are several energy intervals where $G$ is strongly suppressed. The size of such intervals and the $G$ suppression grow with increasing $n_V$.  
Interedge backscattering processes are induced by disorder due to the hybridization of the  edge states with the randomly distributed vacancy-induced localized states in the ribbon.
For weak disorder, $G/G_0 < 1$ only in narrow energy windows that depend on the disorder configuration. 
This mechanism has already been shown to destroy topological protection for such narrow nanoribbons \cite{vannucci2020conductance} and for wide ones in a tight-binding toy model  with much larger vacancy concentrations, $n_V \simeq 2\%$ \cite{Tiwari_2019}.

Let us now address the case of wide nanoribbons, namely, $w \gg \xi$.
Figures \ref{all_transm}(i) to \ref{all_transm}(l) correspond to $w_{110}$ nanoribbons with $n_V $ ranging from $0.13\,\%$ to $0.48\,\%$. 
As in the previous cases, $G$ decreases with increasing $n_V$ for energies outside $\Delta_{\rm TG}$. In distinction, the edge states are topologically protected by time-reversal symmetry and $G= G_0$ over most of the bulk band gap energies. Although topological protection is more robust than in the previous cases, when $n_V \agt 0.30\,\%$, 
$G/G_0$ is strongly suppressed for certain energy intervals within $\Delta_{\rm TG}$, see Figs.~\ref{all_transm}(k) and (l), indicating the presence of vacancy-induced inter-edge backscattering processes.

These results suggest that the number of vacancy-induced states necessary for an effective inter-edge hybridization of the system edge states increases with the nanoribbon width $w$. 
For different building block assemblies we also observe a similar qualitative behavior, namely, a perfect conductance $G_0$ over most of the topological gap with a strong suppression on narrow energy intervals as a function of vacancy concentration. \section{Conclusions and Outlook}
\label{sec:conclusions}

In this paper we have studied the robustness of the conductance quantization against vacancy disorder in large scale nanoscopic buckled bismuthene nanoribbons at the QSH phase.

We have found that vacancies in bismuthene give rise to non-magnetic mid-gap localized states, ruling out local magnetic moments as a mechanism to destroy topological protection as suggested by Ref. 
\cite{Novelli2019}. We have shown that these vacancy-induced mid-gap states can give rise to inter-edge scattering processes. Elastic backscattering is enabled when such states are close in energy and the defect concentration is sufficiently large so that the vacancy-induced states overlap, creating an inter-edge backscattering channel. The backscattering processes depend on the edge state penetration depth, vacancy concentration, and nanoribbon width. The interplay of these quantities has been qualitatively discussed in QSH tight-binding models \cite{Tiwari_2019} and within DFT for narrow systems \cite{vannucci2020conductance}. Here, we have established the presence of vacancy-induced inter-edge backscattering processes in bismuthene nanoribbons of realistic sizes using ab initio techniques.


Our calculations show different transport behavior for bulk and edge states, the first demonstrating localization effects and the latter showing robust topological response for low vacancy concentrations. 
At moderate $n_V$ values, topological protection is destroyed even for wide ribbons $w \gg \xi$.

Our findings are also applicable to other materials in the QSH regime.
Since the penetration depth is material dependent, we conclude that it is possible to engineer different samples with QSH electronic transport behavior by a suitable tuning of the vacancy concentration. Our findings suggest an interesting application for a spintronic device, the level of doping or the width of the device modulates the edge states degeneracy, therefore providing the ON/OFF states for a transistor switch. This might be obtained by changing the chemical potentials on different leads or by applying a gate voltage. In particular, for the topological $w_{65}$ and $w_{110}$ nanoribbons, characterized by a pronounced drop in the conductance becoming broader for energies around the valence band as the level of vacancy concentration increases, allowing the possibility of reaching high ON/OFF ratios. The mechanism to modify the conduction in these nanoribbons does not rely on changing the topology of its band structure as it would happen by applying an electric or magnetic field. 

\begin{acknowledgments}
This work is partially supported by the Coordenação de Aperfeiçoamento de Pessoal de Nível Superior - Brasil (CAPES) - Finance Code 001; FAPESP (Grants 19/04527-0, 16/14011-2, 17/18139-6, and 17/02317-2);
CNPq (Grant 308801/2015-6), INCT-Nanocarbono, and FAPERJ (Grant E-26/202.882/2018). The authors acknowledge the Brazilian National Scientific Computing Laboratory (LNCC) for computational resources of the SDumont computer.
\end{acknowledgments}

\bibliographystyle{apsrev4-2}
\bibliography{main,magnetic_moments}

\end{document}